\documentclass{article}
\pdfoutput=1
\usepackage[a4paper, portrait, margin=0.9in]{geometry}
\usepackage[english]{babel}
\usepackage[utf8]{inputenc}
\usepackage[T1]{fontenc}
\usepackage{helvet}
\usepackage{float}
\usepackage{etoolbox}
\usepackage{graphicx}
\usepackage{titlesec}
\usepackage{caption}
\usepackage{booktabs}
\usepackage{xcolor} 
\usepackage{siunitx}
\usepackage[colorlinks, citecolor = blue, urlcolor = blue]{hyperref}
\usepackage{caption}
\graphicspath{ {./images/} }
\usepackage{scrextend}
\usepackage{fancyhdr}
\usepackage{graphicx}
\newcounter{lemma}

\newcounter{theorem}

\usepackage{cite}
\fancypagestyle{plain}{
	\fancyhf{}

}

\makeatletter
\patchcmd{\@maketitle}{\LARGE \@title}{\fontsize{16}{19.2}\selectfont\@title}{}{}
\makeatother

\usepackage{authblk}

\newsavebox\affbox
\author[1]{\textbf{Jan N. Kirchhof}}
\author[1]{\textbf{Yuefeng Yu}}
\author[1]{\textbf{Denis Yagodkin}}
\author[1,2]{\textbf{Nele Stetzuhn}}
\author[1]{\textbf{Daniel B. de Araújo}}
\author[3]{\textbf{Kostas Kanellopulos}}
\author[4,5]{\textbf{Samuel Manas-Valero}}

\author[4]{\textbf{Eugenio Coronado}}
\author[5]{\textbf{Herre van der Zant}}
\author[1]{\textbf{Stephanie Reich}}
\author[3]{\textbf{Silvan Schmid}}
\author[1]{\textbf{Kirill I. Bolotin}}
\affil[1]{Department of Physics, Freie Universität Berlin, Arnimallee 14, 14195 Berlin, Germany
}

\affil[2]{Max-Born-Institut für Nichtlineare Optik und Kurzzeitspektroskopie, Max-Born-Straße 2A, 12489 Berlin, Germany
}

\affil[3]{Institute of Sensor and Actuator Systems, TU Wien, Gusshausstrasse 27-29, 1040, Vienna, Austria
}
\affil[4]{Instituto de Ciencia Molecular, Universidad de Valencia, Calle Catedrático José Beltrán 2, 46980 Paterna, Spain}
\affil[5]{Kavli Institute of Nanoscience, Delft University of Technology, Lorentzweg 1, Delft, 2628 CJ, The Netherlands}

\titlespacing\section{0pt}{12pt plus 4pt minus 2pt}{0pt plus 2pt minus 2pt}
\titlespacing\subsection{12pt}{12pt plus 4pt minus 2pt}{0pt plus 2pt minus 2pt}
\titlespacing\subsubsection{12pt}{12pt plus 4pt minus 2pt}{0pt plus 2pt minus 2pt}

\titleformat{\section}{\normalfont\fontsize{10}{15}\bfseries}{\thesection.}{1em}{}
\titleformat{\subsection}{\normalfont\fontsize{10}{15}\bfseries}{\thesubsection.}{1em}{}
\titleformat{\subsubsection}{\normalfont\fontsize{10}{15}\bfseries}{\thesubsubsection.}{1em}{}

\titleformat{\author}{\normalfont\fontsize{10}{15}\bfseries}{\thesection}{1em}{}

\title{\textbf{\huge Nanomechanical absorption spectroscopy of 2D materials with femtowatt sensitivity }\\
	}
\date{}    

\begin{document}

\pagestyle{headings}	
\newpage
\setcounter{page}{1}
\renewcommand{\thepage}{\arabic{page}}
\renewcommand{\baselinestretch}{1.50}\normalsize   
\captionsetup[figure]{labelfont={bf},labelformat={default},labelsep=period,name={Figure }}	\captionsetup[table]{labelfont={bf},labelformat={default},labelsep=period,name={Table }}
\setlength{\parskip}{0.5em}
	
\maketitle
\noindent\rule{6.5in}{0.7pt}
	\begin{abstract}
Nanomechanical spectroscopy (NMS) is a recently developed approach to determine optical absorption spectra of nanoscale materials via mechanical measurements. It is based on measuring changes in the resonance frequency of a membrane resonator vs. the photon energy of incoming light. This method is a direct measurement of absorption, which has practical advantages compared to common optical spectroscopy approaches. In the case of two-dimensional (2D) materials, NMS overcomes limitations inherent to conventional optical methods, such as the complications associated with measurements at high magnetic fields and low temperatures. In this work, we develop a protocol for NMS of 2D materials that yields two orders of magnitude improved sensitivity compared to previous approaches, while being simpler to use. To this end, we use electrical sample actuation, which simplifies the experiment and provides a reliable calibration for greater accuracy. Additionally, the use of low-stress silicon nitride membranes as our substrate reduces the noise-equivalent power to $NEP$ = $\SI{890}{fW/\sqrt{Hz}}$, comparable to commercial semiconductor photodetectors. We use our approach to spectroscopically characterize a two-dimensional transition metal dichalcogenide (WS$_2$), a layered magnetic semiconductor (CrPS$_4$), and a plasmonic supercrystal consisting of gold nanoparticles.
  \end{abstract}

\noindent\rule{6.5in}{0.7pt}
\newpage
\section{Introduction}

Nanomechanical resonators emerged as sensitive probes for minuscule forces\cite{force1,weber2016force,Eichler_2022} and atomic-scale masses.\cite{mass111,mass2222,lemme2020nanoelectromechanical,mass1,mass2} By incorporating 2-dimensional (2D) materials into nanomechanical resonators, the miniaturization of these devices has been pushed to the ultimate limit of atomic thickness. Along with this comes a massively reduced effective mass, increased resonance frequencies, easily accessible non-linearity, and the ability to tune resonance frequencies.\cite{steeneken2021dynamics} This technological boost allows using such resonators as sensors for light,\cite{gr-bolometer} magnetic fields,\cite{magnetic-field,jiang2020exchange} sound\cite{verbiest2018detecting,sound1,sound2,lemme2020nanoelectromechanical}, gases\cite{squeeze,lemme2020nanoelectromechanical} or even to study live bacteria.\cite{bacteria}\\
Recently, the use of 2D materials-based resonators as fast and broadband optical spectrometers has been demonstrated.\cite{NMSPEC} In this nanomechanical spectroscopy approach (NMS), changes of the mechanical resonance frequency of a freely suspended 2D material are measured as a function of the illumination photon energy $E_\gamma$. From this, both real and imaginary components of the dielectric function can be extracted. In this measurement the material effectively acts as its own photodetector, leading to broadband sensitivity (UV-THz) of the approach. Furthermore, unlike classical optical approaches, NMS can distinguish between scattered and absorbed light. Finally, NMS is suitable for nanostructures with dimensions smaller than a micron and is applicable at low temperatures and high magnetic fields. Whilst NMS has many fundamental advantages, it currently lacks the sensitivity provided by state-of-the-art optical approaches.\newline
Here, we reduce the noise equivalent power of NMS by two orders of magnitude, down to $NEP$= $\SI{890}{fW/\sqrt{Hz}}$. At the same time, we simplify the method by using piezoelectric actuation, which allows us to study electrically insulating materials, makes sample loading straightforward and provides a simple, but robust calibration. We use this improved method to spectroscopically characterize a range of 2D structures with high resolution. This includes WS$_2$, a classic binary transition metal dichalcogenide (TMD), CrPS$_4$, a layered magnetic semiconductor (ternary TMD), and a plasmonic meta-structure consisting of gold nanoparticles forming a supercrystal.
\cite{mueller2020,mueller2021}

\section{Results}
\subsection{Sample design}
\begin{figure}[H]
	\centering
	\includegraphics[trim=0 70 0 0,clip,width=\textwidth]{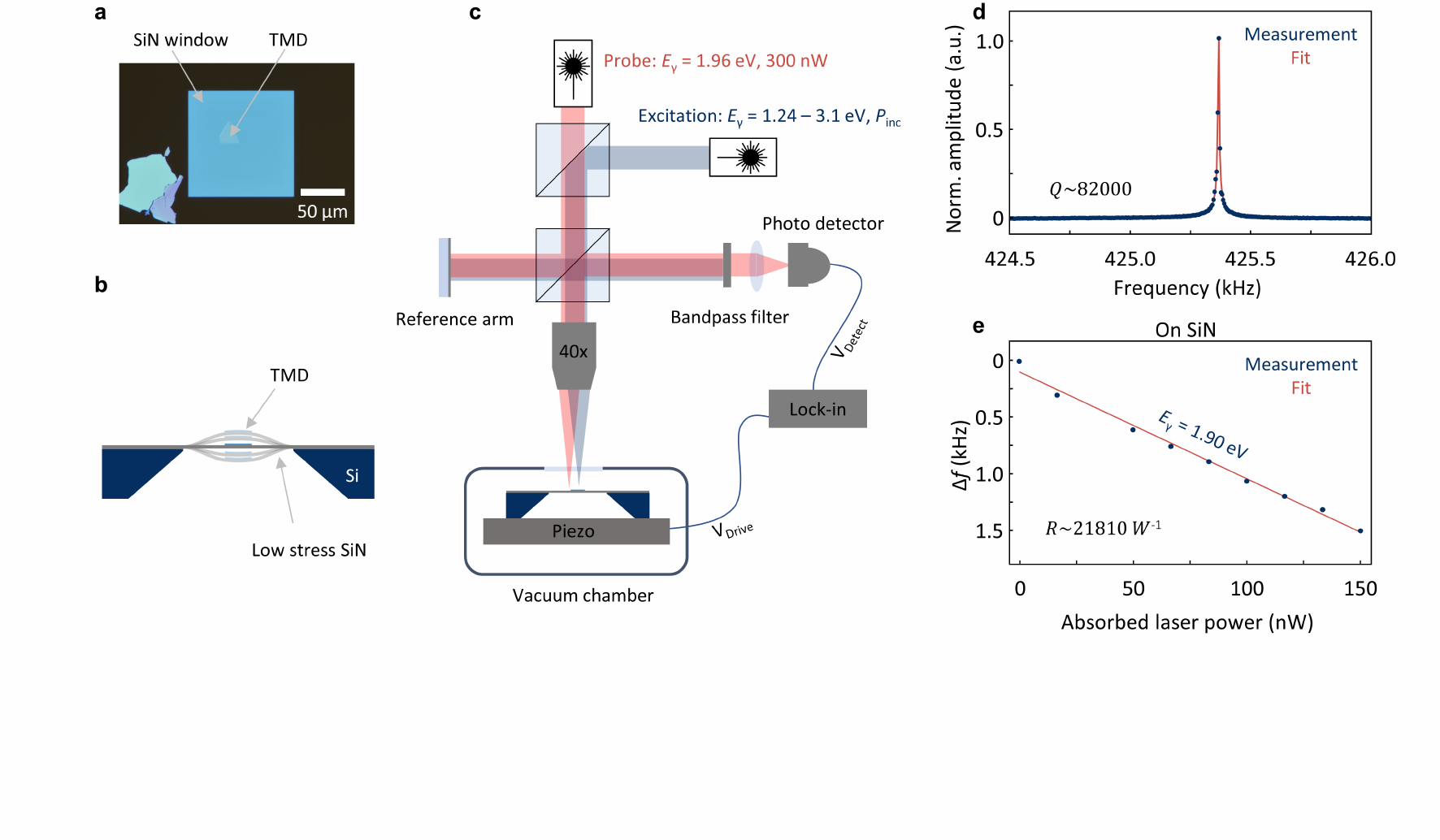}
	\caption{\textbf{SiN-TMD hybrid devices and interferometric motion detection.} \textbf{a}, Optical micrograph of a trilayer WS$_2$ transferd on the center of a low-stress SiN membrane \textbf{b}, Sketch of the SiN-TMD hybrid device. The entire SiN membrane moves out-of-plane and due to the low thermal conductivity of SiN, the TMD flake is thermally decoupled from its environment. \textbf{c}, Motion detection using a Michelson interferometer (red laser) with an additional broadly tunable laser (blue) to heat the resonator via the TMD. The sample is placed in a vacuum  chamber ($p\approx $ ) and the motion of the suspended area is actuated mechanically via a piezoelectric element below the sample. \textbf{d}, Measured amplitude vs. frequency for the device shown in (a). The fundamental mode shows a quality factor of $Q \sim$ 82000. \textbf{e}, Response of the fundamental mode to laser heating (excitation focused on SiN area). We extract a responsivity of  $R_{100\%}$=21810 $W^{-1}$.
}
	\label{fig1}
\end{figure}

At the core of our device is a nanostructure or 2D material of interest that we place on top of a silicon nitride (SiN) membrane, thereby forming a hybrid resonator.\cite{NMSPEC,Gr+SiN,HYBRID} We use SiN for its excellent mechanical properties and record high quality factors ($Q$).\cite{SingleMolecules,sinQ0.bib,sinq1,sinQ2,sinq3} The device is illuminated by a light source of tunable photon energy ($E_\gamma=$ 1.2 to $\SI{3.1}{eV}$). The absorption of light by the 2D material heats up the entire hybrid device. Due to thermal expansion, the in-plane tensile stress within the membrane is released and the resonance frequency downshifts. This downshift in frequency is proportional to the amount of
absorbed laser power ($P_{\text{abs}}(E_{\gamma})$) and allows us to perform absorption spectroscopy.
To understand the measurement's underlying mechanics and optimize our sample design, we start by looking at the fundamental resonance of a square pre-stressed SiN membrane. The resonance frequency is given by $f_0=\frac{1}{L}\sqrt{\frac{\sigma_0}{2\rho}}$, where $L$ is the length, $\sigma_0$ the in-plane tensile stress and $\rho$ the density. When the incoming light heats the material, the stress is reduced by $\Delta\sigma$=$\alpha\Delta T\frac{E}{1-\nu}$, where $\alpha$ is the thermal expansion coefficient, $\Delta T$ is the average temperature increase, $E$ is the Young's modulus and $\nu$ is the Poisson's ratio. Here, $\Delta T$ is directly proportional to $P_{abs}$. The resulting frequency shift $\Delta f=f_0 -f$ can be approximated as $\Delta f$ $\approx f_0\frac{\Delta \sigma}{2\sigma_0}$.\cite{SingleMolecules,gr-bolometer} Thus, in order to maximise the frequency response to laser heating (responsivity) and ultimately improve the sensitivity of NMS, we aim to minimize the in-plane tensile stress $\sigma_0$ within our membranes.\cite{SingleMolecules} 
To do so, we choose SiN membranes ($L$=$\SI{120}{\micro\meter}$, thickness $h$= $\SI{50}{nm}$) grown by low-pressure chemical vapour deposition. By using a silicon-rich stoichiometry, we obtain membranes with low built-in stress. Thin layers of amorphous SiN also have a reduced thermal conductivity of $\kappa \approx \SI{3}{Wm^{-1}K^{-1}}$, \cite{Thermal_cond_SiN} which is beneficial for our experiments as it increases the temperature rise within our device in response to laser heating.\cite{lowstress} To complete the hybrid resonators, we transfer a 2D material of interest onto the SiN-membrane using the PDMS dry transfer technique \cite{PDMS-transfer}. Here, we ensure that we place the 2D material in the centre of the membrane such that there are no thermal links to the rest of the substrate. An optical micrograph of a sample with a 3-layer WS$_2$ is shown in Fig. \ref{fig1}a and and schematically in a side view in Fig.\ref{fig1}b.\newline

The membrane's motion is actuated mechanically by a piezoelectric element below the sample and detected interferometrically using a HeNe laser ($E_{\gamma}=$ $\SI{1.96}{eV}$) with a probe power of $\SI{300}{nW}$ (red in Fig. \ref{fig1}c). A second wavelength-tunable excitation laser (blue in Fig. 1), allows us to controllably heat the resonator, whilst we monitor the changes in resonance frequency. 
Upon sweeping the actuation frequency we find a pronounced fundamental mode at $f_0$ = $\SI{425.367}{kHz}$ as shown in Fig. \ref{fig1}d. From fitting a driven harmonic oscillator response to the experimental data, we estimate (the exact determination requires ring-down measurements) a quality factor of $Q$=82000, comparable to previous results on similar samples.\cite{SingleMolecules} 
Knowing $f_0$, $L$ and  $\rho$= $\SI{3000}{\meter^{3}\kilo\gram^{-1}}$ \cite{sinq1,SingleMolecules}, we calculate the stress in our membrane to be $\sigma_0 \approx$  $\SI{15.6}{MPa}$ -- much lower than commercially available SiN membranes (250 to $\SI{1000}{MPa}$). All measurements are carried out at room temperature (stabilized) and at a pressure of $p \approx \SI{1e-5}{mbar}$. Next, we determine the responsivity of the resonator to absorbed light  $R_{100\%}=\frac{\Delta f}{f_0 P_{\text{abs}}}$. To this end, we focus the probe and excitation beam (set to $E_{\gamma}=$ $\SI{1.9}{eV}$) on the silicon nitride area close to the centre of the sample and vary the laser power of the excitation laser from 0 to $\SI{30}{\micro\watt}$. We use the known absorption ($Abs$=0.5$\%$) of SiN at that photon energy\cite{SingleMolecules} to convert the incident laser power to absorbed laser power ($P_{\text{abs}}=P_{\text{inc}}Abs$) and plot $\Delta f$ vs. $P_{\text{abs}}$ in Fig. \ref{fig1}e. From a linear fit (red), we extract a responsivity of $R_{100\%}$=21810 $W^{-1}$. We will use this quantity for the calibration of our measurements and to calculate the sensitivity of NMS. The measured $R_{100\%}$ is much higher than in previous approaches ($R_{100\%}$=180 $W^{-1}$). \cite{NMSPEC} At the same time, the $Q$ of the improved system is more than an order of magnitude higher than in previous measurements (82000 vs. 5000). Overall, we now have our material of interest implemented into a high quality mechanical hybrid resonator, that is engineered to strongly react to absorbed light.

\subsection{Nanomechanical absorption measurements}
\begin{figure}[H]
	\centering
	\includegraphics[trim=0 140 0 0,clip,width=\textwidth]{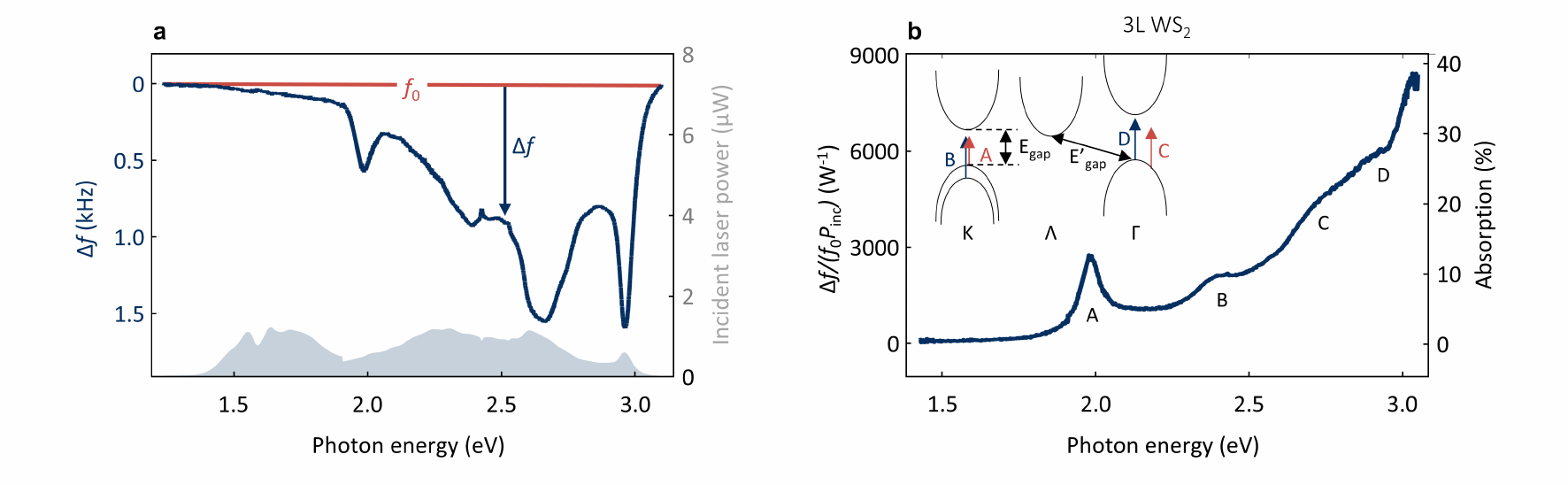}
	\caption{\textbf{Nanomechanical absorption spectroscopy in WS$_2$.} \textbf{a}, Raw frequency response (blue, left y-axis) of the WS$_2$ device as a function of photon energy. The corresponding incident laser power is plotted in grey (right y-axis). Multiple features are visible and towards higher energies, the frequency shift increases as the absorption by the TMD increases. \textbf{b}, Relative responsivity $R=\frac{\Delta f}{f_0 P_{\text{inc}}}$ vs. photon energy. This signal is directly proportional to the absorption coefficient of the TMD and we convert it to absolute units using $R_{100\%}$ (obtained from Fig. 1e). The curve shows clear excitonic features at the expected photon energies. In the inset, we sketch the corresponding transitions for few-layer WS$_2$. 
}
	\label{fig2}
\end{figure}

After optimizing our mechanical system as a sensor for detecting light, we use it to perform absorption spectroscopy on several candidates, starting with the 2D semiconductor WS$_2$. To do so, we track the frequency of the fundamental resonance, whilst varying the photon energy of the excitation laser. In this measurement, the excitation laser is focused on the TMD area and the frequency is measured using a phase-locked-loop (PLL) (details in SI). In Fig. \ref{fig2}a, we plot the frequency shift $\Delta f$ (blue, left y-axis) and the corresponding incoming laser power $P_{\text{inc}}$ (grey, right y-axis) vs. the photon energy of the excitation laser. Upon dividing the frequency shift by the laser power and normalizing it by $f_0$, we obtain the relative responsivity $R(E_\gamma)=\frac{\Delta f(E_\gamma)}{f_0 P_{\text{inc}}(E_\gamma)}$ (Fig. \ref{fig2}b). This signal is directly proportional to the absorption coefficient of the 2D material. Finally, we use the beforehand determined $R_{100\%}$ to calculate the absorption coefficient of the TDM in absolute units: $Abs(E_\gamma)=R(E_\gamma)/R_{100\%}$ - as plotted in Fig. \ref{fig2}b on the right y-axis. In this spectrum, we find clear peaks (A-D), that correspond to different excitonic species at the expected energies.\cite{ABCD} In the inset we sketch the corresponding transitions for few layer WS$_2$ (following Ref. \cite{ABCD}). We note that while the 2D material absorbs most of the light, some fraction also will reach the SiN below. This could offset our measurement results, but SiN is an amorphous insulator with very weak light-matter interaction, resulting in low absorption ($\sim $0.5$\%$) and a rather flat spectrum in the spectral region of interest. We, therefore neglect the absorption of light in SiN below the TMD.

\subsection{Characterizing exotic 2D materials}
\begin{figure}[H]
	\centering
	\includegraphics[trim=0 140 0 0,clip,width=\textwidth]{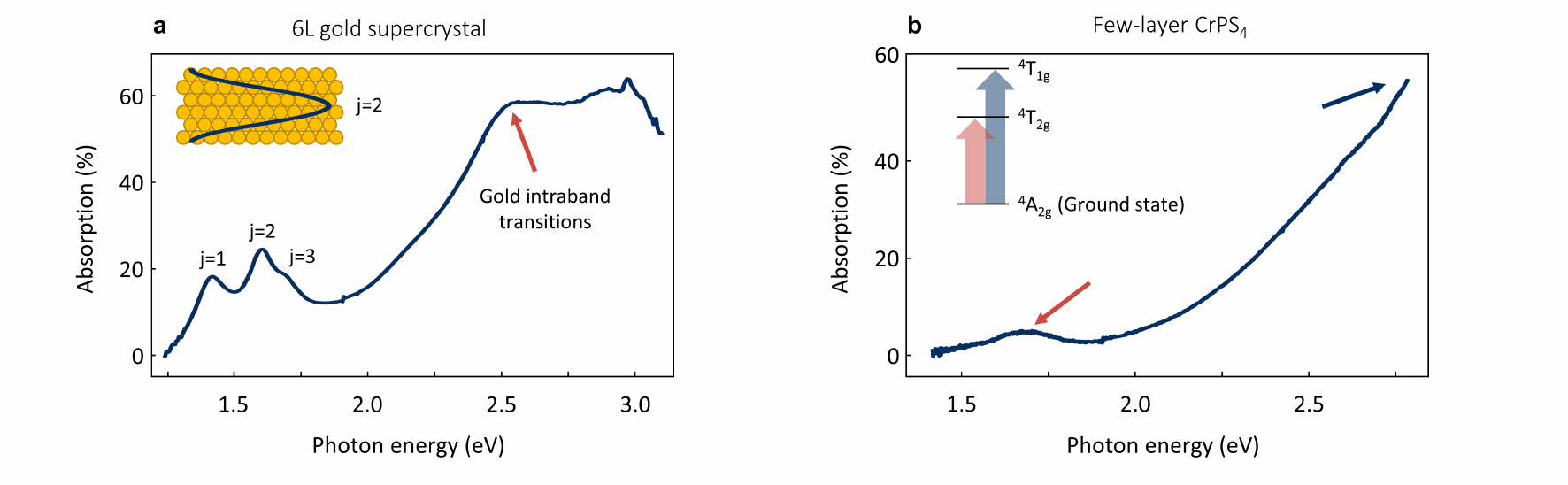}
	\caption{\textbf{Nanomechanical absorption spectroscopy in exotic 2D materials} \textbf{a}, Absorption vs. photon energy for a 6-layer gold plasmonic super crystal (thickness of $\SI{350}{nm}$). In the low energy range up to $\SI{1.8}{eV}$, we find pronounced absorption peaks associated with standing wave plasmon polaritons within the super crystal. Above $\SI{2.5}{eV}$ the light is absorbed by intraband transitions in the gold particles. The inset shows a sketch of the gold nanoparticles (yellow circles) arranged in a supercrystal, in which plasmonic modes form standing waves ($j=2$-mode shown in blue). When excited on resonance these modes absorb light and are visible as peaks in our measurement. \textbf{b}, Absorption vs. photon energy for a few-layer ($\sim 5$) 2D antiferromagnet CrPS$_4$. This material is a semiconductor with a broad absorption edge around $\SI{1.68}{eV}$, which we observe in our measurements. 
}
	\label{fig3}
\end{figure}

Next, we extend our method to study more exotic 2D materials. We choose two materials that are particularly suited for our method and for which it is expected to produce advantages. The first one is a layered crystal (supercrystal) made from plasmonic nanoparticles. These gold nanoparticles (diameter $\SI{56}{nm}$) are embedded into a polymer matrix forming a closely spaced fcc supercrystal.\cite{mueller2020,mueller2021} Such supercrystals have been recently shown to enter the deep strong light-matter coupling regime, in which light within a material can no longer be seen as a perturbation to the properties of the material.\cite{mueller2020,mueller2021} Instead, the material properties are almost entirely determined by light-matter interaction. \cite{mueller2020,mueller2021}. We grow the supercrystals by self-assembly on a liquid-liquid interface (details inf Ref. \cite{mueller2020,mueller2021,liquid-liquid}), carefully transfer them onto the SiN membranes by the PDMS dry transfer technique\cite{PDMS-transfer} and perform our nanomechanical absorption spectroscopy measurements. In Fig. \ref{fig3}a we show the absorption spectrum obtained for a 6-layer supercrystal. Starting from low photon energies, we find three peaks associated with plasmonic modes within the supercrystal (j=1..3 comp. inset Fig. \ref{fig3}). As we go to higher energies, the absorption increases towards the intraband transitions of gold and plateaus around $\SI{2.5}{eV}$. Interestingly, NMS works excellently despite the supercrystal being much thicker ($\SI{350}{nm}$ vs. $\SI{2}{nm}$) and also much heavier than the TMD. Plasmonic structures are known to scatter a significant amount of light that cannot be distinguished from absorbed light by common optical measurement methods. In contrast, NMS is only sensitive to absorbed light and therefore ideal to study plasmonic systems. 

Our second choice is a 2D antiferromagnet CrPS$_4$ with a Néel temperature for the bulk material of $T_{\text{Neel}}$= $\SI{36}{K}$.\cite{magnet1} The study of 2D magnetism requires measurements at low temperatures and high magnetic fields, which is challenging for conventional optics but is easier using NMS. We exfoliate and transfer a few-layer (approx. 5) thick flake of CrPS$_4$ onto a SiN membrane and perform NMS (see Fig. \ref{fig3}b). In the spectrum, we find a broad peak around $\SI{1.68}{eV}$. This peak belongs to the d$-$d transition of the Cr$^{3+}$ ions from the $^4$A$_{\text{2g}}$ to the $^4$T$_{\text{2g}}$ state. Towards higher photon energies the absorption increases, as we approach higher-order transitions (e.g. $^4$A$_{\text{2g}}$ to $^4$T$_{\text{1g}}$). This spectrum measured above the antiferromagnet's Néel temperature is expected and in line with literature reports. \cite{magnet1,magnet2,magnet3}. Thin CrPS$_4$ is rather sensitive to photodamage \cite{magnet1,magnet3}, which can be problematic for classical optical approaches. For NMS however small laser powers ($P_{\text{inc}}\approx$$\SI{1}{\micro\watt}$) are sufficient to induce sizeable frequency shifts (comp. Fig.\ref{fig2}a), which can help preserve the quality of sensitive samples. Overall, we have spectroscopically characterized a range of exotic 2D materials with high resolution and believe that these measurements highlight the broad applicability of NMS.

\subsection{Sensitivity}
\begin{figure}[H]
	\centering
	\includegraphics[trim=0 140 0 0,clip,width=\textwidth]{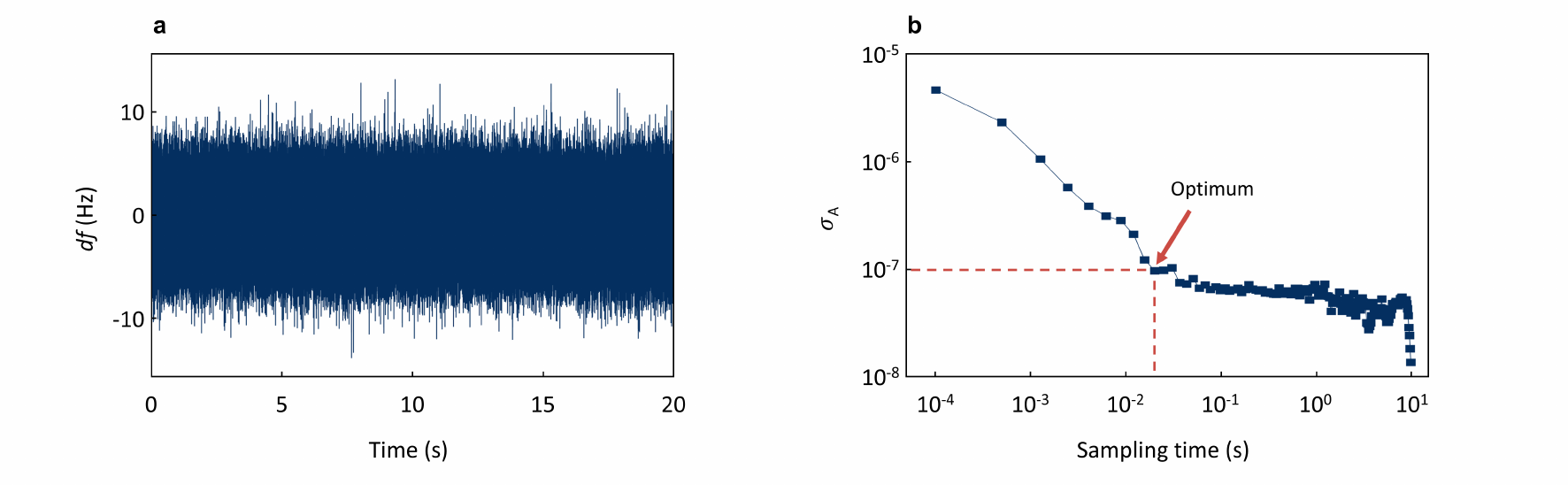}
	\caption{\textbf{Sensitivity} \textbf{a}, Stability measurement of frequency vs. time without laser illumination, measured using a phase-locked loop. \textbf{b}, Allan deviation $\sigma_{A}$ vs. sampling time in log-log scale derived from a). For an optimal
sampling time of $\SI{20}{ms}$ (red arrow), we extract a noise equivalent power of $NEP$= $\SI{890}{fW/\sqrt{Hz}}$)
. 
}
	\label{fig4}
\end{figure}

In order to compare NMS to state-of-the-art optical approaches, we determine its sensitivity (noise equivalent power $NEP$) and response time ($\tau$). In our measurements, we use frequency shifts to probe the amount of absorbed light. In order to quantify the noise in our measurements, we, therefore, look at frequency fluctuations, which we then convert into power noise using $R_{100\%}$. \cite{NMSPEC,gr-bolometer,IR_spec_schmid,schmid-thz-trampoline,schmid-thz-1}. Assuming white noise, the noise equivalent power ($NEP$) can be derived from the fractional frequency noise power spectral density ($S_y(0)$) and is given by:\cite{nep1,NEP2}  
\begin{eqnarray}
	NEP=\frac{\sqrt{S_y(0)}}{R_{100\%}}=\frac{\sigma_{\text{A}}\sqrt{2t_{\text{sampling}}}}{R_{100\%}}
	\label{Eq1}
\end{eqnarray} 
where $\sigma_\text{A}$ is the Allan deviation of the frequency measurement\cite{ALLAN} and $t_{\text{sampling}}$ is the sampling time. To obtain $\sigma_\text{A}$, we perform a frequency stability measurement in the PLL configuration with the excitation laser turned off (Fig. \ref{fig4}a). From this data, we derive $\sigma_\text{A}$ vs. sampling time (Fig. \ref{fig4}b). We choose an optimal value of $\sigma_\text{A}$= $\SI{9.7e-8}{}$ for $t_{\text{sampling}}$ = $\SI{20}{ms}$ and using $R_{100\%}=\SI{21810}{W^{-1}}$, we obtain $NEP$= $\SI{890}{fW/\sqrt{Hz}}$. This value is two orders of magnitude lower than in our previous approach ($NEP$=  $\SI{90}{pW/\sqrt{Hz}}$) \cite{NMSPEC}. \newline 
The sensitivity of NMS is now comparable to commercially available avalanche photodetectors (APDs) for the same spectral range with $NEP$=  $\SI{200}{fW/\sqrt{Hz}}$ (Thorlabs APD130A(/M)). APDs are highly sensitive, but also overload quickly. Compared to these commercial devices, NMS-based devices show a higher dynamic range ($\SI{84}{dB}$ vs. $\SI{69}{dB}$ - both for $\SI{1}{s}$ integration time) and can easily detect hundreds of µW (details in SI).

To assess the measurement speed of NMS, we simulate time-dependent laser heating in our sample and extract a response time for our mechanical system of $\tau=$ $\SI{800}{\micro\second}$ (simulations in SI), which is in line with experimental data on similar devices.\cite{schmid-thz-trampoline,tau} The fast response time, allows us to sweep the excitation energy rapidly and we obtain the $\Delta f$ vs. $E_{\gamma}$ traces presented above in a matter of seconds. 

\section{Discussion}
We presented a simplified and improved method of nanomechanical absorption spectroscopy. With a sensitivity of $NEP$= $\SI{890}{fW/\sqrt{Hz}}$ and a response time $\tau=$$\SI{800}{\micro\second}$, the method now is a promising alternative to classical optical approaches, whilst it overcomes long-standing limitations. At the same time, we show that using SiN as a reference material provides a robust and straightforward calibration. We demonstrated the broad applicability of NMS by spectroscopically characterizing a 2D semiconductor, a layered plasmonic supercrystal and a novel 2D antiferromagnet. The key points for the improvement of the measurement sensitivity are stress reduction in the SiN membranes and thermal decoupling of the 2D material, which leads to enhanced responsivity and improved sensitivity. To further reduce $NEP$, we could use temperature regulation of the sample to controllably minimize the tensile stress $\sigma_0$. Upon cooling the sample, the silicon frame shrinks more than the suspended SiN membrane, due to their difference in thermal expansion coefficient. This reduces the stress in the suspended SiN and would allow measuring very close to $\sigma_0=0$, which would mean further increased responsivity and thus improved sensitivity. We note that in comparison to our previous experiments, the 2D material is not suspended in this work but in direct contact with the SiN. This changes the dielectric environment of the 2D material and can affect excitons and other quasiparticles in 2D materials. We find that this has a large impact on photoluminescence  measurements, where it completely suppresses the emission, but does not affect absorption spectra. \cite{dielectric_func_tmds} \newline 
We also aim to characterize 2D materials at low temperatures using NMS. In the limit of $T \rightarrow  $$\SI{0}{K}$, the specific heat of any material goes to zero and along with it the thermal conductivity $\kappa$ and thermal expansion coefficient $\alpha$. If $\alpha$ decreases faster than $\kappa$, our method will not work anymore, because the responsivity is proportional to  $\frac{\alpha}{\kappa}$.\cite{SingleMolecules} To exclude this scenario, we do preliminary measurements of the responsivity of a bare SiN membrane as a function of temperature. For this, we use a membrane with higher pre-stress ($\sigma_0= $$\SI{240}{MPa}$) to avoid buckling upon cooldown and find that even at $\SI{4.2}{K}$, the membrane shows considerable responsivity (see SI). This paves the way for future experiments at low temperatures and will allow the unlocking of exciting physics in a large range of 2D materials.

\section{Data availability statement}
The data that support the findings of this study are available upon reasonable request from the authors.
\section{Acknowledgement }
This work was supported by Deutsche Forschungsgemeinschaft (DFG, German Research Foundation, Project ID 449506295, 328545488 and 504656879), CRC/TRR 227 (Project B08), ERC Starting Grant No. 639739, CSC 202006150013 and the Consolidator Grant DarkSERS (772108). Samuel Mañas-Valero thanks the Generalitat Valenciana for a postdoctoral fellow APOSTD-CIAPOS2021/215.
\bibliographystyle{iopart-num}

\bibliography{mainSI} 

\newpage

\section{Supplementary information}
\setcounter{figure}{0} 

\subsection{Phase locked-loop measurements}
All resonance frequency measurements were performed using a lock-in amplifier MFLI from Zurich Instruments. To set up the phase-locked-loop (PLL), we first sweep the drive frequency and then use this frequency and phase information to lock onto the resonance using the built-in PLL functionality of the lock-in amplifier. We choose a PLL bandwidth of $\SI{1}{kHz}$ and a resolution bandwidth of $\SI{5}{kHz}$. To improve the stability and speed of the PLL, we use the lock-in amplifier’s PLL advisor to optimize the feedback parameters. For the frequency stability measurements (Fig. \ref{fig4}a), we adjust the drive voltage just below the onset of the non-linearities and try to avoid external noise sources. We use a temperature controller to stabilize the sample temperature at $T=\SI{300}{K}$. To account for the remaining thermal drift in our measurements, we perform a linear correction on the measured frequency traces.
\subsection{Finite element method (FEM) simulations}
To simulate laser heating in our samples, we use the structural mechanics and heat transfer module of Comsol Multiphysics (version 6.0). The device geometry including the under-laying silicon frame is shown in Supplementary Fig. 1a. Next, we implement a  Gaussian heat source centrally on the TMD ($P_{abs}$=$\SI{10}{\micro\watt}$, $r_{laser}=\SI{500}{nm}$) and calculate the resulting temperature profile (Supplementary Fig. 1b). To estimate the saturation power of our detector, we increase the amount of absorbed laser power until we reach the damage threshold of the TMD, which we expect at an average temperature of $T_{max}\approx\SI{550}{K}$ in the TMD. According to our simulations, this value is reached at $P_{abs}=\SI{145}{\micro\watt}$. To calculate the dynamic range, we divide the maximum power by the minimum detectable power ($P_{min}=\eta\sqrt{BW}$) for an integration time of $\SI{1}{s}$ ($BW=\sqrt{\SI{1}{Hz}}$). We calculate a ratio of $\SI{2.3e{8}}{}$, which expressed in decibel yields $\SI{84}{dB}$.

\begin{figure}[H]
	\centering
 \renewcommand{\figurename}{Supplementary Figure}

	\includegraphics[trim=0 10 0 0,clip,width=\textwidth]{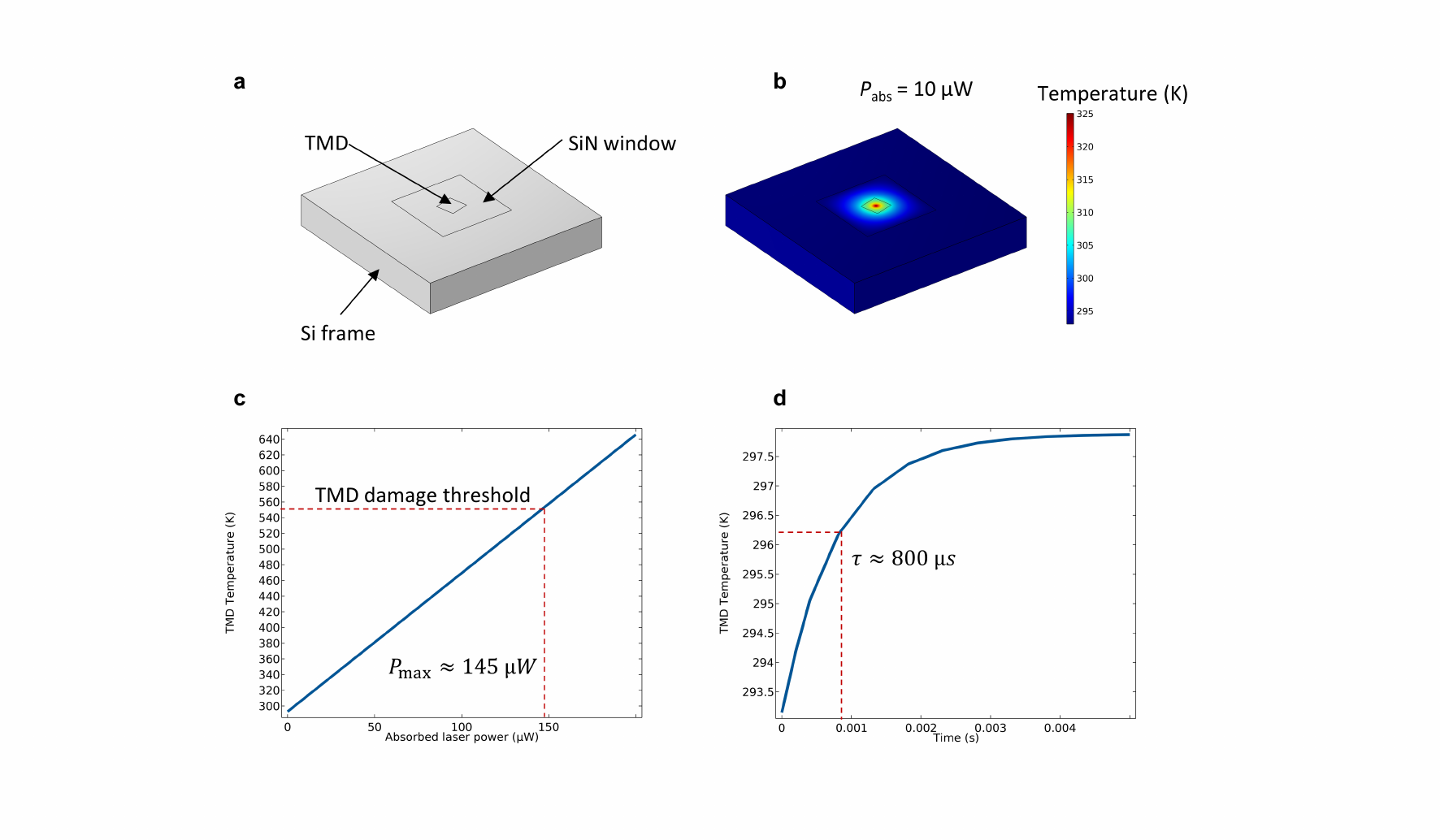}
	\caption{\textbf{FEM Simulations} \textbf{a}, Sketch of the device geometry \textbf{b}, Temperature profile upon laser illumination \textbf{c}, Resulting average temperature in the 3-layer WS$_2$ sample.  \textbf{d} Simulated time response of the average temperature in the TMD as laser heating is introduced at t = 0. We extract a response time of $\tau=$ $\SI{800}{\micro\second}$. To determine the thermal response time, we simulate the average temperature of the TMD vs. time after turning on the illumination (Supplementary Fig. 1d). We extract a response time of $\tau\approx$ $\SI{800}{\micro\second}$. 
}
	\label{figS1}
\end{figure}

\subsection{Temperature depended responsivity}
We aim to extend our NMS measurements to low temperatures. Upon cooling, the thermal constants of the SiN membrane change, which will affect the responsivity ($R_{100\%}\sim$ $ \frac{\alpha}{\kappa}$). To ensure that our measurements will work at low temperatures we perform preliminary tests with a bare SiN membrane (Norcada NX5150A). The membrane has higher built-in stress ($\sigma_0=$$\SI{240}{MPa}$) to avoid buckling when the sample is cooled. In Supplementary Fig. 2, we show the measured responsivity as a function of cryostat temperature. At liquid helium temperature, we find a responsivity comparable to the room temperature value. We conclude that NMS will work over a large range of temperatures.
\begin{figure}[H]
	\centering
 \renewcommand{\figurename}{Supplementary Figure}
	\includegraphics[trim=0 40 0 0,clip,width=\textwidth]{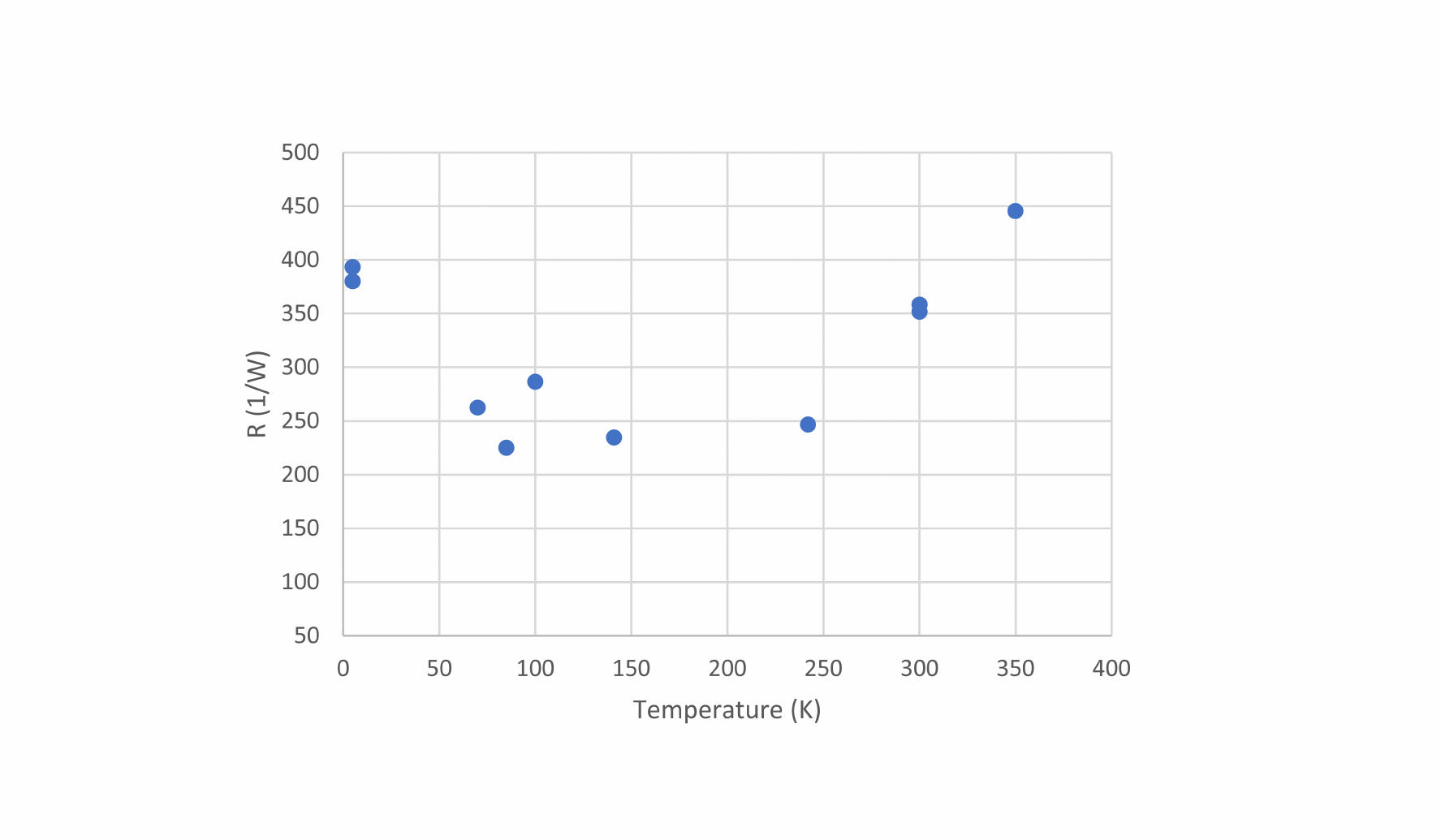}
	\caption{\textbf{Responsivity vs. device temperature} \textbf{a}, Responsivity of a SiN membrane (with $\sigma_0=\SI{240}{MPa}$) as a function of cryostat temperature. 
}
	\label{figS2}
\end{figure}

\end{document}